\documentclass[
    twocolumn,
    11pt,
	prd,
	amssymb,
	preprintnumbers,
	secnumarabic,
	nofootinbib,
	superscriptaddress
	]{article}

\pdfoutput=1
\usepackage{graphicx}
\usepackage{enumitem}
\usepackage{latexsym}
\usepackage{amsfonts}
\usepackage{amssymb}
\usepackage{color}
\usepackage{amsmath}
\usepackage{slashed}
\usepackage{dcolumn}
\usepackage{verbatim}
\usepackage{float}
\usepackage{xspace}
\usepackage[normalem]{ulem}
\usepackage{imakeidx}
\usepackage{lipsum, babel}
\makeindex[columns=3,title=Index 1: Mythos]
\makeindex[columns=3,title=Index 2: Physics,name=indphy]
\usepackage[
pdfauthor={Nirmal Raj}]{hyperref}

\definecolor{mygreen}{rgb}{0.3,0.6,0.3}
\definecolor{myblue}{rgb}{0.3,0.3,0.7}

\hypersetup{
  pdfauthor={Nirmal Raj},
  urlcolor=mygreen,
  colorlinks=true,
    linkcolor=mygreen
}

\allowdisplaybreaks

\begin{document}

\title{Mythos in Fundamental Physics}

\author{Nirmal Raj}


\maketitle

\section*{Introduction}


There must be something about the myths of ancient Greece that stirs physicists, which must be why they draw upon them heavily for naming experiments, machines, codes, and phenomena.
The practice is particularly ubiquitous in fundamental physics, perhaps because one of the chief functions of Greek mythology parallels the chief goal of the field: to address the mysteries of Nature.  
Some of the best myths are ``etiological" tales relating the origin of seasons, lightning, disease, echoes, the peculiar specks of light moving in the sky, the pleasing pattern of spider webs, and so forth.
Throw in the dense, fascinating network of storylines and the capacity of mythos to tap deep into human nature (prompting Joseph Campbell to declare that ``myths are public dreams, dreams are private myths"), and you can see why mythology is generous in lending names to 
planets,
moons,
stars, 
collections of stars, 
space missions,
months, 
days of the week, 
chemical elements, and 
biological species.
In this article we will look at the names fundamental physics has borrowed from Greek myth.\\

A remarkable thing about the body of stories that is Greek mythology is that nearly all the characters come from a single family, so that one way to advance much of the timeline is to simply build the family tree.
That is how I have presented the material here.
To keep matters brief, I have had to excise numerous narrative strands, and omit much about the characters that give them colour and depth.
I hope that the few central threads I have selected make the reader curious to find out more; a bibliography is appended.
I have also provided two indices, one themed on myth and the other on physics. \\

And thus we begin, with the helium-electron-photon\index[indphy]{helium, electron, photon}\footnote{Incidentally, helium was named for  Helios\index{Helios}, Titan god of the Sun, as it was was discovered in the solar spectrum;
``electron" has its roots in a Greek word for amber, which also supplied the name of Elektra\index{Elektra}, nephew of Menelaus (whom we will meet here);
and ``photon" has a root that also branched out Phoebus Apollo.} -- the $\alpha \beta \gamma$ -- of Greek mythos, the 
\clearpage

\section*{Immortals}

\subsection*{First Order}

Before there was a before, there only was Chaos\index{Chaos}, either a state of emptiness or a jumbled heap of formless raw material, depending on whom you ask. Out of Chaos came\\

NYX\index{Nyx}\index[indphy]{Phenomena!Nyx} [\href{https://www.nature.com/articles/s41550-020-1131-2}{\em a nearby stellar stream}], who personified the night, and\\

GAIA\index{Gaia}\index[indphy]{Space probes!GAIA} [\href{https://en.wikipedia.org/wiki/Gaia_(spacecraft)}{\em satellite observatory} {\em out to map a billion stars}], who personified Earth. In some versions she is the mother of\\ 

ECHIDNA\index{Echidna}\index[indphy]{Particle accelerators and detectors!ECHIDNA} [\href{https://www.ansto.gov.au/our-facilities/australian-centre-for-neutron-scattering/neutron-scattering-instruments/echidna-high}{\em diffractometer} for neutron scattering], ``mother of all monsters", especially Chimera, Cerberos, Sphinx, Hydra, and Scylla, all of whom will appear in these pages.
Gaia also mothered \\

PYTHON\index{Python} \index[indphy]{Space probes!Python} [\href{https://inspirehep.net/experiments/1110610}{\em cosmic microwave background telescope}], a serpentine monster at the Earth's centre. Via parthenogenesis Gaia begets Uranus\index{Uranus/Ouranos}, the first sky god, and together they populate the world with the

\subsection*{Second Order}

Among the children of Gaia and Uranus were the\\

CYCLOPS\index{Cyclops}\index[indphy]{Particle accelerators and detectors!CYCLOPS} [\href{https://www.ill.eu/users/instruments/instruments-list/cyclops/description/instrument-layout}{\em CYlindrical CCD Laue Octagonal Photo Scintillator}], three one-eyed giants who would make thunderbolts for their nephew Zeus (see below) in the War of the Titans, bringing us to\\

TITAN\index{Titan}\index[indphy]{Instruments!TITAN} [\href{https://titan.triumf.ca/}{\em TRIUMF's Ion Trap for Atomic and Nuclear science}], name of twelve of Gaia-Uranus' offspring (and their descendants), original gods of the world, among whom was\\

HYPERION\index{Hyperion}\index[indphy]{Particle accelerators and detectors!Hyperion} \index[indphy]{Phenomena!HYPERION} [\href{https://inspirehep.net/literature/1517573}{\em gamma-ray detector array} and \href{https://inspirehep.net/literature/2656659}{\em HYPerluminous quasars at the Epoch of ReionizatION}, {\em quasar sample}] -- who personified the sun along with his son Helios -- and his wife\\

THEIA\index{Theia}\index[indphy]{Astroparticle detectors!THEIA} [\href{https://inspirehep.net/literature/1764050}{\em a planned large-scale neutrino experiment} {\em detecting both Cerenkov and scintillator light}], the goddess of sight, mother of\\
 
EOS\index{Eos}\index[indphy]{Computation!EOS}\index[indphy]{Astroparticle detectors!EOS} [\href{https://eos.github.io/doc/index.html}{\em software framework for flavor physics} and \href{https://nino.lbl.gov/eos/detector.html}{\em prototype to THEIA}], goddess of the dawn, whose nephew was\\

IRIS\index{Iris}\index[indphy]{Instruments!IRIS} [\href{https://fiveyearplan.triumf.ca/teams-tools/iris-isac-charged-particle-reaction-spectroscopy-station/}{\em rare isotope spectroscopy} {\em at TRIUMF}], goddess of the rainbow, and whose cousin was\\

DORIS\index{Doris}\index[indphy]{Particle accelerators and detectors!DORIS} [\href{https://www.desy.de/research/facilities__projects/doris/index_eng.html}{\em storage ring and accelerator} {\em at DESY}], meaning ``gift", a sea goddess, who had another cousin,\\

PROMETHEUS\index{Prometheus}\index[indphy]{Computation!Prometheus} [\href{https://arxiv.org/abs/2304.14526}{\em neutrino telescope simulation}], the best friend of Zeus (see below) at first, but for a while punished as the darkest enemy for handing mortals the secret of fire. Prometheus had a brother,\\

ATLAS\index{Atlas}\index[indphy]{Particle accelerators and detectors!ATLAS} [\href{https://atlas.cern/}{\em A Toroidal LHC Apparatus}, {\em co-discoverer of, well, the god particle}], whose punishment at the end of the War of the Titans, meted out by Zeus (see below), is to hold up the sky. The War came about when his aunt Rhea coupled with\\

CRONUS\index{Cronus/Cronos/Kronos; see also Saturn}\index[indphy]{Phenomena!chronon} [\href{https://en.wikipedia.org/wiki/Chronon}{\em chronon}: {\em a hypothetical quantum of time}], the last of the 12 Titans. 
The Romans, who imported Greek mythology wholesale, called him\\

SATURN\index{Saturn; see also Cronus}\index[indphy]{Particle accelerators and detectors!SATURN} [\href{https://arxiv.org/abs/1512.06198}{\em Scintillator And Tape Using Radioactive Nuclei}]. 
Envious of his father Uranus' power, he conspires with his mother Gaia to overthrow Uranus and appoint himself ruler, but not before incurring a curse from the tormented father: ``You shall suffer this hurt by {\em your} child". In response, he swallows his first five babies with Rhea. The sixth is switched for a linen-swaddled stone; this boy grows up in secret in Crete, and returns to overthrow Cronus, in the process causing him to vomit the five children in his belly, now fully grown. The six siblings, with help from their uncles the Cyclops and a few cousins, win the fierce War of the Titans that follows, and establish the final and

 \subsection*{Third Order}
 
OLYMPUS\index{Olympus}\index[indphy]{Particle accelerators and detectors!OLYMPUS} [\href{https://www.desy.de/research/facilities__projects/olympus/index_eng.html}{{\em detector at DORIS}}], the mountain that was the new headquarters of the ``\href{https://www.theoi.com/greek-mythology/olympian-gods.html}{Olympian gods}". As with the Titans, they were (eventually) a dozen. Their queen was Rhea's eldest (and the last expelled by Cronus),\\ 

HERA\index{Hera; see also Juno}\index[indphy]{Particle accelerators and detectors!HERA} [\href{https://www.desy.de/research/facilities__projects/hera/index_eng.html}{\em Hadron-Electron Ring Accelerator} {\em at DESY}], goddess of marriage and family, known in Roman myth as\\

JUNO\index{Juno; see also Hera}\index[indphy]{Astroparticle detectors!JUNO} [{\em the imminent} \href{https://en.wikipedia.org/wiki/Jiangmen_Underground_Neutrino_Observatory}{\em Jiangmen Underground Neutrino Observatory}, {\em incidentally also a lepton-hadron scattering experiment}]. Her husband was 
\\

ZEUS\index{Zeus; see also Jupiter}\index[indphy]{Particle accelerators and detectors!ZEUS} 
\index[indphy]{Computation!Zeus21}[\href{https://en.wikipedia.org/wiki/ZEUS_(particle_detector)}{\em detector at HERA}] and 
[\href{https://inspirehep.net/literature/2633073}{\em Zippy Early Universe Solver}, {\em package for computing 21-cm-line signal}], Roman name Jupiter\index{Jupiter; see also Zeus}, the sixth fruit of the union of Cronus and Rhea, he who had liberated the other five, Olympian god of the sky and thunder, king of the gods, the All-Father. As we shall see, he sired untold characters in Greek mythos, Olympian gods among them, one of whom was\\

HERMES\index{Hermes; see also Mercury}\index[indphy]{Particle accelerators and detectors!HERMES}\index[indphy]{Space probes!HERMES} [\href{https://en.wikipedia.org/wiki/HERMES_experiment}{\em another experiment at HERA}, and the \href{https://ui.adsabs.harvard.edu/abs/2012MNRAS.424.1614O/abstract}{\em Herschel Multi-tiered Extragalactic Survey}], Roman name Mercury\index{Mercury; see also Hermes}, the god of thieves and tricksters. His speed earned him the office of the gods' messenger [{\em NASA's Mercury probe MESSENGER\index[indphy]{Space probes!MESSENGER} was used to}  \href{https://arxiv.org/abs/2006.10008}{\em measure the neutron lifetime}]. His best friend was his half-brother by Zeus, Apollo\index{Apollo, or Phoebus Apollo}, sometimes called Phoebus Apollo (``Apollo the bright"), the Olympian god of (among other things) the sun and prophecies. The latter he delivered at his temple in\\
 
DELPHI\index{Delphi}\index[indphy]{Particle accelerators and detectors!DELPHI}\index[indphy]{Computation!DELPHES} [\href{https://en.wikipedia.org/wiki/DELPHI_experiment}{\em DEtector with Lepton, Photon and Hadron Identification} {\em at LEP} and \href{https://cp3.irmp.ucl.ac.be/projects/delphes}{\em DELPHES, detector response simulation code}], via the high priestess by name of\\

PYTHIA\index{Pythia}\index[indphy]{Computation!PYTHIA} [\href{https://pythia.org/}{\em Monte Carlo event generator}], who served as oracle. While Apollo handled the truth, it was\\ 

ATHENA\index{Athena; see also Minerva}\index[indphy]{Space probes!ATHENA} \index[indphy]{Particle accelerators and detectors!ATHENA} [\href{https://www.cosmos.esa.int/web/athena/about-athena}{\em future X-ray satellite} to study, e.g., black holes, and  \href{https://inspirehep.net/literature/2164827}{\em A Totally Hermetic Electron Nucleus Apparatus} for the Electron-Ion Collider], known to the Romans as\\ 

MINERVA\index{Minerva; see also Athena}\index[indphy]{Particle accelerators and detectors!MINER$\nu$A} [\href{https://minerva.fnal.gov/}{\em Main Injector Experiment for $\nu$-A,} {\em neutrino experiment at Fermilab}], that was the Olympian goddess of wisdom. Also the goddess of war, she carried a shield, the\\

AEGIS\index{Aegis}\index[indphy]{Particle accelerators and detectors!AEgIS} [\href{https://en.wikipedia.org/wiki/AEgIS_experiment}{\em Antimatter Experiment: Gravity, Interferometry, Spectroscopy}], in which the head of Medusa (see below) would later be emblazoned. Fathered by Zeus and mothered by\\

METIS\index{Metis}\index[indphy]{Space probes!METIS} [\href{https://ipa.phys.ethz.ch/research/ResearchProjects/e-elt-metis0.html}{\em Mid-infrared ELT Imager and Spectrograph}], his tutor, Athena takes a vow of chastity, much like her half-sister\\

DIANA\index{Diana; see also Artemis}\index[indphy]{Computation!DIANA} [\href{https://diana-hep.org/}{\em software toolkit for colliders}], Roman for Artemis\index{Artemis; see also Diana}, the Olympian goddess of hunting and the moon, and Apollo's twin. 
Artemis and Apollo were born to Leto on the island Delos, which was originally Leto's sister\\

ASTERIA\index{Asteria}\index[indphy]{Computation!Asteria} [\href{https://inspirehep.net/literature/2693522}{\em tool to compute capture of particle dark matter in celestial bodies}], who transformed herself into a wandering piece of land to escape Zeus' randy advances.
Another kind of immortal fathered by Zeus was the\\

MUSE\index{Muse}\index[indphy]{Particle accelerators and detectors!MUSE} [{\em a proposed} \href{https://inspirehep.net/literature/1925378}{\em MUon Scattering Experiment}], who inspired creativity, born as they were to Mnemosyne\index{Mnemosyne}, the Titan goddess of memory. The Muses were nine, of which\\
 
CALLIOPE\index{Calliope}\index[indphy]{Instruments!CALLIOPE} [\href{https://inspirehep.net/literature/227524}{\em multi-particle magnetic spectrometer}], mother of Orpheus (see below), was the Muse of epic poetry,\\
 
EUTERPE\index{Euterpe}\index[indphy]{Particle accelerators and detectors!EUTERPE} [\href{https://aip.scitation.org/doi/10.1063/1.1143023}{\em electron storage ring}] of music,\\

URANIA\index{Urania}\index[indphy]{Instruments!u-RANIA}\index{Instruments!Urania} [\href{https://inspirehep.net/literature/2037490}{\em argon extraction plant} {\em for DarkSide-20k} and {\em future }\href{https://inspirehep.net/literature/1796235}{\em neutron imaging detector}] of astronomy,\\

CLIO\index{Clio}\index[indphy]{Instruments!CLIO} [\href{https://inspirehep.net/literature/1452852}{\em free electron laser}] of history, and\\

ERATO\index{Erato}\index[indphy]{Computation!ERATO} [\href{https://inspirehep.net/literature/423233}{\em event generator} {\em for LEP}] of literature, science, and the arts.\\ 

Besides Hermes, Apollo, Artemis, and Athena, three other Olympian gods were born to Zeus: one, Hephaestus\index{Hephaestus; see also Vulcan}\index{Vulcan; see also Hephaestus} (in Rome, Vulcan), god of the forge and fire; two, Ares (Mars)\index{Ares; see also Mars}\index{Mars; see also Ares}, the war god and father of \\

PHOBOS\index{Phobos}\index[indphy]{Particle accelerators and detectors!PHOBOS} [\href{https://inspirehep.net/experiments/1108662}{\em detector at RHIC}], god of fear -- it is Ares' name that lends the epithet\\

AREIA\index{Areia}\index[indphy]{Instruments!Aria} [\href{https://inspirehep.net/literature/1842194}{\em argon distillation plant} {\em for DarkSide-20k}], ``war-like", to the Olympians Athena, Artemis and Aphrodite (see below) -- and three, Dionysus (Bacchus)\index{Dionysus; see also Bacchus}\index{Bacchus; see also Dionysus}, god of wine and merry-making, whom Zeus himself bore in his thigh, hence his name meaning ``twice born". By his gift of wine, Dionysus once caused the death of an Athenian, whose maiden daughter Erigone\index{Erigone; see also Virgo} then hanged herself in grief. Out of pity, Zeus put her in the sky as the constellation\\
 
VIRGO\index{Virgo; see also Erigone}\index[indphy]{Space probes!VIRGO} [\href{https://en.wikipedia.org/wiki/Virgo_interferometer}{\em gravitational wave interferometer}]. 
Only Hephaestus and Ares were Hera's; the queen's indignations and spiteful jealousies are a mythological staple, in one instance giving the name of a Greek hero (into whom we will run), and in another causing the death of her servant, the many-eyed giant\\

ARGUS\index{Argus}\index[indphy]{Particle accelerators and detectors!ARGUS} [\href{https://en.wikipedia.org/wiki/ARGUS_(experiment)}{\em A Russian-German-United States-Swedish collaboration}, {\em detector at DORIS}]. Now Zeus had an aunt via Uranus,\\ 

VENUS\index{Venus; see also Aphrodite}\index[indphy]{Particle accelerators and detectors!VENUS} [\href{https://research.kek.jp/people/odaka/HEP/VENUS/index.html}{\em magnetic spectrometer} {\em at the $e^+e^-$ collider TRISTAN}], Olympian goddess of love and beauty. In Greece she was Aphrodite\index{Aphrodite; see also Venus}, ``from the foam", and we would like to tell you why she was so named, but this is a family website. Her son,\\

EROS\index{Eros; see also Cupid}\index[indphy]{Space probes!EROS} [\href{http://eros.in2p3.fr/}{\em gravitational microlensing experiment}], or Roman-tically\\

CUPID\index{Cupid; see also Eros}\index[indphy]{Astroparticle detectors!CUPID} [\href{https://cupid-mo.mit.edu/}{\em CUORE Upgrade with Particle IDentification}, {\em experiment looking for neutrinoless double beta decay}], was notorious for carrying wayward arrows that ignited passion in their marks. Rounding up the Olympian pantheon are the other four disgorged by Cronus: Poseidon\index{Poseidon; see also Neptune}, Roman name Neptune\index{Neptune; see also Poseidon}, god of the ocean, earthquakes and horses;\\ 

HESTIA\index{Hestia; see also Vesta}\index[indphy]{Computation!HESTIA} [\href{https://arxiv.org/abs/2008.04926}{\em  High-resolutions Environmental Simulations of The Immediate Area}, cosmological simulations of the Local Group], Roman name Vesta\index{Vesta; see also Hestia}, goddess of home and hearth, who willingly steps down to make room for Dionysus;\\

CERES\index{Ceres; see also Demeter}\index[indphy]{Particle accelerators and detectors!CERES} [\href{https://inspirehep.net/literature/563457}{\em lead-gold colliding experiment} {\em at CERN SPS}], Roman for Demeter\index{Demeter; see also Ceres}, goddess of farmers and fertility; and\\

HADES\index{Hades; see also Pluto}\index[indphy]{Particle accelerators and detectors!HADES} [\href{https://inspirehep.net/literature/813810}{\em High-Acceptance Dielectron Spectrometer} {\em used to search for dark photons}], Roman name\\ 

PLUTO\index{Pluto; see also Hades}\index[indphy]{Particle accelerators and detectors!PLUTO} [\href{https://en.wikipedia.org/wiki/PLUTO_detector}{\em detector at DORIS}], king of the underworld, where the dead go. Keeping him company down there are his wife Persephone\index{Persephone/Cora/Kore/Proserpina}, daughter of Demeter, and\\

CHARON\index{Charon}\index[indphy]{Computation!$\chi$aro$\nu$} [\href{https://inspirehep.net/literature/1809475}{\em software package} {\em for neutrino fluxes sourced by dark matter}], who ferries dead souls across the subterranean river \\

STYX\index{Styx}\index[indphy]{Particle accelerators and detectors!STYX} [\href{https://inis.iaea.org/search/search.aspx?orig_q=RN:47024695}{\em Straw Tube Young student eXperiment} {\em built from the decommissioned ZEUS}], boundary between Earth and Underworld; and \\

CERBERUS\index{Cerberus/Kerberos}\index[indphy]{Particle accelerators and detectors!CERBEROS} [\href{https://inspirehep.net/literature/1309237}{\em beam detector} {\em at GSI}], three-headed serpent-tailed dog at the gates of Hades that keeps the dead from leaving.\\

That Hades is able to populate his once-barren kingdom at all is due to an idea that came to Zeus as he fretted restlessly in the peaceful but boring order that he himself had brought about. The idea was to create as play-things conscious, appeasing, worshipping creatures in the image of the gods themselves, only... mortal. 

\clearpage

\section*{Mortals}

Numerous mortals sought to defy the gods, garnering their fury and our acclaim. We begin our survey of them -- the {\em Alpher-Bethe-Gamow of Greek mythos}\index[indphy]{Alpher, Bethe, Gamow}, if you will -- with those famous for their exploits in the underworld. \\

ORPHEUS\index{Orpheus}\index[indphy]{Astroparticle detectors!ORPHEUS} [\href{https://inspirehep.net/literature/531235}{\em dark matter experiment}, and \href{https://inspirehep.net/literature/1285702}{\em proposal} {\em to detect axion dark matter}], son of Apollo and the muse Calliope (see above), was the finest musician that ever was. Widowed abruptly, he goes down to Hades, charms its keepers with songs of love, and returns to Earth with the spirit of his wife Eurydice\index{Eurydice} -- almost. (Orpheus will return in these pages as an Argonaut.) Then there's\\

MINOS\index{Minos}\index[indphy]{Particle accelerators and detectors!MINOS}\index[indphy]{Computation!MINOS} [\href{https://en.wikipedia.org/wiki/MINOS}{\em Main injector neutrino oscillation search} {\em at Fermilab}, and \href{https://github.com/joelwwalker/AEACuS}{\em Machine Intelligent Optimization of Statistics}], a Cretan king who tormented Theseus and the Minotaur (see below for both), whose father of the same name is appointed in Hades a judge of the dead with the deciding vote, the other two arbiters being\\

AECUS\index{Aecus}\index[indphy]{Computation!AECUS} \& RHADAMANTHUS\index{Rhadamanthus}\index[indphy]{Computation!RHADAMANTHUS} [\href{https://github.com/joelwwalker/AEACuS}{\em Algorithmic Event Arbiter and Cut Selector \& Recursively Heuristic Analysis, Display, and Manipulation: The Histogram Utility Suite}]. It was probably with their sanction that eternal punishments were meted out to the following three.\\

IXION\index{Ixion}\index[indphy]{Instruments!IXION} [\href{https://aip.scitation.org/doi/10.1063/1.1706312}{\em rotating plasma device}], a king who made passes at Hera, doomed to spin forever in a wheel of fire; he will re-appear here in the thread on the hero Jason;\\

TANTALUS\index{Tantalus}\index[indphy]{Particle accelerators and detectors!TANTALUS} [\href{https://www.researchgate.net/publication/281059137_Tantalus_the_First_Dedicated_Synchrotron_Radiation_Source}{\em synchrotron radiation source}], who invites the Olympian pantheon over for a banquet, but gets it into him to test their omniscience by serving up his own (chopped up and boiled) son; he now stands in Hades waist-high in a pool of water that recedes whenever he stoops in thirst, and under the branches of a fruit tree that retreats whenever he reaches in hunger. As for his son, the gods resurrect him, this Pelops\index{Pelops}, who will go on to grandfather Menelaus, a key player in the Trojan War (coming up below);\\

SISYPHUS\index{Sisyphus}\index[indphy]{Phenomena!Sisyphus cooling} [\href{https://en.wikipedia.org/wiki/Sisyphus_cooling}{\em laser cooling technique}], a wily ruler who twice cheats Death.
The hubris is noted; when after a long and happy life he goes at last down under, he is consigned to roll a boulder up a steep gradient in an infinite loop: should he succeed in reaching the summit, eternal freedom from Hades would be his, but the rock is designed to slip near the top. Another victim of such illusion of choice was\\

PANDORA\index{Pandora}\index[indphy]{Computation!PANDORA} [\href{https://arxiv.org/abs/1308.4537}{{\em a particle flow algorithm}}], the first mortal woman, made in Hephaestus' forge. Her name means ``all-gifted" (c.f. Doris above), as she was given a quality each by an Olympian god. Unable to contain her curiosity, she opens a jar (not a box) given her by Zeus as a wedding gift under instructions to {\em never} do so; the jar was full of evils. Consequences for humanity until today echo. Speaking of which --\\

ECHO\index{Echo}\index[indphy]{Instruments!ECHO} [\href{https://www.kip.uni-heidelberg.de/echo/}{{\em Electron Capture Holmium experiment}}], a talkative nymph of the mountains who crossed goddess Hera by protecting Zeus, and thereby incurred a curse: her only conversation would be repetition of the final words heard from another. Her misery is ended when Aphrodite frees her of her body, leaving behind just her voice that even now dwells in canyons and empty rooms. A similar afterlife awaited\\

DAPHNE\index{Daphne}\index[indphy]{Particle accelerators and detectors!DAFNE} [\href{https://en.wikipedia.org/wiki/DAFNE}{\em Double Annular $\Phi$ Factory for Nice Experiments}, {\em $e^+e^-$ collider at INFN National Laboratory of Frascati}], a freshwater nymph who attracted the unwanted attention of the sun god Apollo, leading to a chase that only ended when her river god father turned her into a laurel tree. These tales may be compared with the transformation of\\ 

ADONIS\index{Adonis}\index[indphy]{Particle accelerators and detectors!ADONE} [\href{https://en.wikipedia.org/wiki/ADONE}{\em ADONE}, {\em harder version of Anello Di Accumulazione $e^+e^-$ collider}], lover of Aphrodite and Persephone, to anemones wherever his handsome blood fell whilst dying from the wounds inflicted by a wild boar, and with that of\\
   
  ARACHNE\index{Arachne}\index[indphy]{Computation!ARACHNE} [\href{https://inspirehep.net/literature/946964}{\em web-based event display} {\em at MINER$\nu$A}], who outweaved Minerva/Athena, the goddess of wisdom, with tapestries depicting scenes of Zeus' infidelities, and for this double insolence was punished by the goddess with strikes to the head. The humiliation drove Arachne to the noose, but Athena in grudging admiration turned her into the very first spider so that she and her descendants may continue to weave perfect designs.
  The gods bestowed immortality in many such ways, their favourite being ``catasterizing" -- placing among the stars, usually as a constellation. For instance,\\
  
 AURIGA\index{Erichthonius; see also Auriga}\index{Auriga; see also Erichthonius}\index[indphy]{Space probes!AURIGA} [\href{https://en.wikipedia.org/wiki/AURIGA}{\em Antenna Ultracriogenica Risonante per l'Indagine Gravitazionale Astronomica}, {\em resonant bar gravitational wave detector}], the Charioteer constellation, was once King Erichthonius of Athens.
 Born to the Olympian god Hephaestus and primordial goddess Gaia, and raised by Athena (who was the intended mother; how this accident came about is {\em not} left to the imagination in the original myth), he had impressed Zeus enough with his chariot-driving talents to be recognized thus.
 Likewise,\\

 CYGNUS \index{Cycnus/Cygnus} \index[indphy]{Astroparticle detectors!CYGNUS} [\href{https://inspirehep.net/literature/1813839}{\em proposed gaseous dark matter detector}] was originally Cycnus turned into the swan constellation as relief from his mourning and sorrow over the death of his lover, Phaethon -- Helios' son, More famously,\\

ORION\index{Orion}\index[indphy]{Particle accelerators and detectors!ORION} [\href{https://www-project.slac.stanford.edu/orion/}{\em accelerator and beam research collaboration at SLAC}], a giant hunter, was raised to the heavens by a sorry Artemis after killing him by mistake. She was tricked by her brother Apollo, jealous of her love for fellow hunter Orion, for whom she eschewed her vow of chastity. This was after Apollo convinced Gaia to finish Orion with a giant scorpion, which failed, but got a constellation of its own in the zodiac. Just like\\

ARIES\index{Aries}\index[indphy]{Particle accelerators and detectors!ARIES}
[\href{https://aries.web.cern.ch/}{\em Accelerator Research and Innovation for European Science and Society} ], a gold-shaded flying ram that helped rescue Phrixus\index{Phrixus}, the step-son of Ixion (see above), from a sacrificial end. It takes him to Colchis\index{Colchis}, where it is promptly sacrificed, and the king of Colchis gifted its fleece. Years later, the hero Jason is tasked to retrieve the Golden Fleece by undertaking a most dangerous voyage, which he does on\\

ARGO\index{Argo, Argonaut}\index[indphy]{Astroparticle detectors!ARGO} [\href{http://webusers.fis.uniroma3.it/~nucleare/argo/argo_data.html}{\em cosmic ray detector} and \href{https://ui.adsabs.harvard.edu/abs/2019APS..APRH17002W/abstract}{\em future dark matter detector}], a ship built for speed, not to be confused with Argos, the dog of Odysseus (see below). Its crew was an assembly of the most intrepid heroes alive at the time: once an Argonaut, always an \\

ARGONAUT\index[indphy]{Particle accelerators and detectors!ARGONAUT}\index{Tantalus}\index[indphy]{Astroparticle detectors!ArgoNeuT} [\href{https://inspirehep.net/literature/139224}{\em 1978 neutrino detector} and \href{https://www.symmetrymagazine.org/article/august-2008/bonnie-and-the-argoneuts}{\em ArgoNeuT} {\em at Fermilab}]. Braving countless adventures, the Argonauts reach Colchis whence Jason steals the Golden Fleece\index{Golden Fleece} with help from the princess, who is smitten by him, by name of\\

MEDEA\index{Medea}\index[indphy]{Particle accelerators and detectors!MEDEA}
[\href{https://www.sciencedirect.com/science/article/abs/pii/016890029290497R}{\em multi element detector array}], who returns with Jason on the Argo, and helps him escape her father's navy. Among further adventures, the Argonauts encounter the three Sirens\index{Sirens; see also Aglaope, Parthenope, Thelxiope}, whose enchanting songs lure sailors to cannibalistic doom. But the music of Orpheus (see above), on board for just such occasions, saves them. The sirens are often named as Aglaope\index{Aglaope; see also Sirens}, Thelxiope\index{Thelxiope; see also Sirens} and\\

PARTHENOPE\index{Parthenope; see also Sirens}\index[indphy]{Computation!PARTHENOPE} [\href{http://parthenope.na.infn.it/}{\em Public Algorithm Evaluating the Nucleosynthesis of Primordial Elements}]. Another famous encounter with the Sirens, with a different deed of ingenuity to escape their clutches, is made by Odysseus (see below). But to know him, we must first know \\

DANAE\index{Danae}\index[indphy]{Astroparticle detectors!DANAE} [\href{https://inspirehep.net/literature/1749054}{\em imminent dark matter detector} {\em underground}], whose father, fearful of a prophecy that his grandchild would do him in, confines her to a heavily guarded chamber beneath the palace. But the prison -- and Danae -- are not impregnable: Zeus appears as golden rain and begets a child, Perseus\index{Perseus}, one of the first great heroes. As a youth, Perseus (now living in a different land) is tasked by his king to fetch the head of\\
 
MEDUSA\index{Medusa}\index[indphy]{Instruments!MEDUSA} [\href{https://www.geneseo.edu/nuclear/medusa-callibration}{\em an array of 800 proton recoil detectors}], a mortal who, incurring the wrath of Athena for soiling her temple by coupling with Poseidon there, was turned into a snake-haired monster. Make eye contact with her, and you will turn into stone -- be literally petrified. Perseus evades this fate while slicing her head off by the use of a mirror-like shield. From her severed head then emerge her children by Poseidon, the warrior Chrysaor\index{Chrysaor}, and\\

PEGASUS\index{Pegasus}\index[indphy]{Computation!PEGASUS} [\href{https://link.springer.com/article/10.1140/epjc/s10052-020-7898-6}{\em Monte Carlo event generator}], a winged horse, for Poseidon was the creator and god of horses. He was tamed by the hero Bellerophon\index{Bellerophon}, who registers him as a personal vehicle. The two combine forces to slay the \\

CHIMERA\index{Chimera}\index[indphy]{Particle accelerators and detectors!CHIMERA} [\href{https://www.dfa.unict.it/en/content/chimera}{\em Charged Heavy Ion Mass and Energy Resolving Array}], 
an offspring of Echidna (see above) -- a fire-breathing monstrosity with the head and body of a lion, a second head, that of a goat, jutting out of its back, and the tail of a serpent. In some accounts it was Chimera who had given birth to the Nemean Lion\index{Nemean Lion}, slain by \\

HERACLES\index{Heracles; see also Hercules}\index[indphy]{Computation!HERACLES} [\href{https://cds.cern.ch/record/208150}{\em event generator} {\em for HERA}], Roman name Hercules\index{Hercules; see also Heracles},
great-grandson of Perseus and another son of Zeus. Often considered the greatest of Greek heroes, Heracles (``Hera's glory") was so named by his mortal parents to appease the seething Hera. As atonement for accidentally killing his wife and children, Heracles must undertake 12 tasks, the ``Labours of Heracles" -- getting rid of the Nemean Lion was the first. The second was to eliminate the Lernaean\\
 
HYDRA\index{Hydra, or Lernaean Hydra}\index[indphy]{Instruments!HYDRA} [\href{https://www.oeaw.ac.at/smi/research/precision-experiments/hydra}{\em Hydrogen and Deuterium Ramsey/Rabi Apparatus}], yet another of Echidna's litter of horrors. It had multiple heads, each growing back twofold when snipped off, yet Heracles slays it. Following the accomplishment of the twelve Labours, Heracles sacks Troy\index{Troy} to carry out a grudge against its ruler, killing every royal except two: \\

 HESIONE\index{Hesione}\index[indphy]{Computation!HESIONE} [\href{https://aip.scitation.org/doi/abs/10.1063/1.5123031}{\em simulation of beam-target interaction}], a princess whom Heracles' friend Telamon\index{Telamon} loves, and Podarces\index{Podarces; see also Priam}, her ten year old brother whose life she ransoms for a golden veil of Athena.
 Podarces vows to rebuild Troy as the greatest, wealthiest city state in the Greek world, and doesn't mind his new name, meaning ``bought one": Priam\index{Priam; see also Podarces}. Priam keeps his promise, and fathers Paris, who sparks the epic Trojan War\index{Trojan War} that dooms mighty Troy. But perhaps equal or more blame must be ascribed to \\

 ERIS\index{Eris; see also Discordia}\index[indphy]{Computation!ERIS} [\href{https://iopscience.iop.org/article/10.1088/0004-637X/742/2/76}{\em spiral galaxy simulation}], Roman name Discordia\index{Discordia; see also Eris}, goddess of strife, who is understandably the only god not invited to the wedding of Peleus\index{Peleus} and Tethys\index{Tethys}, parents of the hero Achilles (see below). She comes nevertheless in the middle of the feast, held near the cave of \\

CHIRON\index{Chiron}\index[indphy]{Computation!CHIRON} [\href{https://link.springer.com/article/10.1140/epjc/s10052-014-3249-9}{\em package} {\em for chiral perturbation theory}], son of Cronus, an all-wise tutor of several heroes named here, and a\\

CENTAUR\index{Centaur}\index[indphy]{CENTAUR} [\href{https://centaur.tamu.edu/}{\em Center for Excellence in Nuclear Training and University-based Research}], of upper half human and lower half horse fame. Eris then rolls out a golden apple and leaves; on it are inscribed the words ``To the fairest". Hera, Athena, and Aphrodite each claim it, the dispute then taken to the judgement of Paris\index{Paris}, reputed to be impartial. Hera offers him power, Athena wisdom, but Aphrodite he picks the instant she offers him the hand of\\ 

 HELEN\index{Helen, or Helen of Troy}\index[indphy]{Astroparticle detectors!ELENA} [\href{http://serena.iaps.inaf.it/elena.html}{\em Emitted Low-Energy Neutral Atoms} {\em or ELENA, detector on board Mercury Planet Orbiter}], the most beautiful across time, space and mortality, daughter of\\

 LEDA\index{Leda}\index[indphy]{Particle accelerators and detectors!LEDA} [\href{https://www.sciencedirect.com/science/article/abs/pii/S0168900200004794}{\em Louvain–Edinburgh Detector Array} {\em for nuclear beams}], with whom Zeus couples as a swan, causing her to lay an egg from which hatches the girl.\\
 
  Helen is married off to Menelaus\index{Menelaus}, great-grandson of Tantalus (see above) and king of Sparta\index{Sparta}. Paris journeys there, tricks Menelaus into leaving and returns to Troy with Helen; the husband, not taking kindly to the duplicity, assembles a colossal navy of neighboring kingdoms (``the Argives"\index{Argive}) to attack Troy. Meanwhile up in Olympus, Hera and Athena take the side of the Argives, recruiting whichever other god could be influenced. Aphrodite does the same on the side of the Trojans. The Trojan War has begun, providing Homer\index{Homer} with material for the {\em Iliad}\index{Iliad}. Troy is led by\\
 
HECTOR [\href{https://inspirehep.net/literature/755654}{\em beamline simulator}], elder brother to Paris, eventually felled by \\

ACHILLES\index{Achilles}\index[indphy]{Computation!ACHILLES} [\href{https://inspirehep.net/literature/2081799}{\em event generator for lepton-nucleus scattering}], the most glorious of the battle heroes. 
His immortal mother Tethys, distraught by the finiteness of his mortal lifespan, had dipped her infant son in the sacred underworld river Styx (see above) so it may bestow upon him invincibility. 
But she had held him by the heel. 
It is this heel that Paris' poisoned arrow finds near the close of the Trojan war, doing its owner in.
The siege wages on for ten years, ending when the Argives pretend to have left, leaving behind the\\

TROJAN HORSE\index{Trojan horse}\index[indphy]{Computation!Trojan Horse} [\href{https://www.annualreviews.org/doi/abs/10.1146/annurev-nucl-102419-033642}{\em a computational method} {\em for nuclear astrophysics quantities}], a wooden giant made to look like a gift to Athena (who had an important shrine in Troy), but in reality a seamlessly sealed enclosure for Argive warriors. The Horse is rolled into the otherwise impenetrable walls of Troy, whereupon the occupants climb out and unlock the gates. Troy is finished. The Trojan Horse, as are several solutions to seemingly impossible problems, is the cunning of\\

ULYSSES\index{Ulysses; see also Odysseus}\index[indphy]{Computation!ULYSSES} [\href{https://arxiv.org/abs/2007.09150}{\em Universal LeptogeneSiS Equation Solver}], Greek name Odysseus\index{Odysseus; see also Ulysses}, king of Ithaca, who had tried to evade the Trojan draft against a prophecy that it would cost him 20 years before returning home. Sure enough, a vengeful Poseidon ensures that an otherwise straightforward return voyage from Troy becomes a protracted ten-year nightmare of unwelcome adventures, the stuff of Homer's {\em Odyssey}\index{Odyssey}. Among these is a long stop at the island of Aeaea\index{Aeaea}, where lived\\

CIRCE\index{Circe}\index[indphy]{Astroparticle detectors!CIRCE} [\href{https://www.ia.forth.gr/project/451}{\em project} to study ultra-high-energy cosmic rays], a minor goddess seeking to retain Odysseus for love, reluctantly letting him go with advice to visit Hades' underworld for further instructions from the wise. This eventually leads him to encounter, besides the Sirens (see above) and other horrific wonders,\\

SCYLLA\index{Scylla}\index[indphy]{Computation!SCYLLA} [\href{https://www.scylladb.com/users/case-study-cern-optimizes-computing-resources-with-scylla/}{\em open-source database} {\em used by ALICE @ LHC}]
and 

CHARYBDIS\index{Charybdis}\index[indphy]{Computation!CHARYBDIS} [\href{https://inspirehep.net/literature/624071}{\em event generator} {of black holes at hadron colliders}] on either side of a narrow strait, the one a six-headed sailor-devouring monster, the other a whirlpool with personality.
During Odysseus' prolonged absence, his son Telemachus seeks the counsel of wise\\

NESTOR\index{Nestor}\index[indphy]{Astroparticle detectors!NESTOR} [\href{https://inspirehep.net/experiments/1110447}{\em Neutrino Extended Submarine Telescope with Oceanographic Research}], king of Pylos, who is regrettably unhelpful. 
But at long last Odysseus is reunited with his wife,\\

PENELOPE\index{Penelope}\index[indphy]{Computation!PENELOPE} [\href{https://www.oecd-ilibrary.org/nuclear-energy/penelope-a-code-system-for-monte-carlo-simulation-of-electron-and-photon-transport_9d2cc3d5-en}{\em Monte Carlo simulation} {\em of electron and photon transport}], who had been artfully keeping suitors away in her husband's absence. Now you may think that Paris had had no way of knowing how fraught with consequences his act of taking Helen would be, but no, \\

THESEUS\index{Theseus}\index[indphy]{Space probes!THESEUS}\index[indphy]{Particle accelerators and detectors!THESEUS} [\href{https://www.isdc.unige.ch/theseus/}{\em Transient High-Energy Sky and Early Universe Surveyor}, {\em imminent infrared and X-ray satellite for cosmology}, and \href{https://inspirehep.net/experiments/1110544}{\em The Second Experiment Underground at Soudan} {\em for neutrino oscillation measurements}] got there first, and paid for it with his mother Aethra, enslaved by Helen after rescue came. Athens' greatest hero, Theseus slew the Minotaur, creature with a bull's head and a man's body, imprisoned by King Minos (see above) in the heart of a Labyrinth. The Minotaur was mothered by Minos' queen consort\\

PASIPHAE\index{Pasiphae}\index[indphy]{Space probes!PASIPHAE} [\href{https://pasiphae.science/}{\em Polar-Areas Stellar-Imaging in Polarization High-Accuracy Experiment}] and fathered by the Cretan Bull\index{Cretan Bull} (don't ask), whose capture happened to be Heracles' seventh labour, a task Theseus himself undertook afterward in Marathon on the mainland.
Minos had captured Theseus as food for the Minotaur, but the hero secures the help of Minos' daughter,\\

ARIADNE\index{Ariadne}\index[indphy]{Particle accelerators and detectors!ARIADNE}\index[indphy]{Computation!ARIADNE} [\href{https://hep.ph.liv.ac.uk/ariadne/}{\em liquid argon detector} {\em imaging particle tracks}, \href{https://inspirehep.net/literature/1924337}{\em software library} {\em for particle track reconstruction}, and \href{http://www.dnp.fmph.uniba.sk/cernlib/asdoc/ariadne.html}{\em QCD cascade generator}], who hands him a ball of string to track his trajectory. When the king discovers the escape and elopement, he jails in a tower\\

DAEDALUS\index{Daedalus}\index[indphy]{Particle accelerators and detectors!DAE$\delta$ALUS}
 [\href{https://inspirehep.net/literature/882191}{\em neutrino experiment} {\em to measure CP-violating angle, rendered DAE$\delta$ALUS}], designer of the Labyrinth, and his firstborn,\\

ICARUS\index{Icarus}\index[indphy]{Astroparticle detectors!ICARUS} [\href{https://icarus.fnal.gov/}{\em Imaging Cosmic And Rare Underground Signals}, {\em neutrino detector at Fermilab}]. Father and son escape by attaching makeshift wings of bird feathers and wax, with the father issuing strict orders to fly neither too high (to keep the sun from melting the wax) nor too low (to keep the sea from soaking the feathers). Alas, the youthful Icarus cannot help soaring high, and pays the ultimate penalty plunging into the sea. Not unlike the\\

 SPHINX\index{Sphinx}\index[indphy]{Particle accelerators and detectors!SPHINX} [\href{https://arxiv.org/abs/physics/0504035}{\em spectrometer and collaboration}], who, too, was winged, albeit by nature. (The Sphinx must not be confused with the\\
 
 PHOENIX\index{phoenix}\index[indphy]{Particle accelerators and detectors!PHENIX} [\href{https://en.wikipedia.org/wiki/PHENIX_detector}{\em Pioneering High Energy Nuclear Interaction eXperiment} {\em at RHIC}], a bird that resurrects from its predecessor's ashes.) A creature with human head and lion's body, the Sphinx blocked the entrance to Thebes, posing to travellers a riddle {\em Which creature walks on four legs in the morning, two in the afternoon, and three in the evening?} Those failing to answer, which was everyone, became her snack. Her menace is ended by  \\

OEDIPUS\index{Oedipus}\index[indphy]{Computation!OEDIPUS} [\href{https://www.hep.phy.cam.ac.uk/theory/software/oedipus_old.html}{\em Onium Evolution Dipole Interaction and Perturbative Unitarisation Computation}], who solves the riddle ({\em Man! Get out of here!}), sending her on a seaward plunge. But there ended his good fortune, for everything he touched from then on turned to profound tragedy, quite unlike\\

MIDAS\index{Midas}\index[indphy]{Computation!MIDAS} [\href{https://en.wikipedia.org/wiki/Maximum_Integrated_Data_Acquisition_System}{\em Maximum Integrated Data Acquisition System} {\em for particle detectors}], to whom Dionysus granted a wish for golden touch, delightful at first, but which by and by parted Midas from food and family.\\

-----

\clearpage

\onecolumn

\section*{Afterword}

\ \ \ {\bf Future naming}

  I have collected several dozen names from Greek myth that have found a place in fundamental physics. 
  Are there any more that await use? 
  Only several hundreds. 
  The interested experimentalist or coder should have no difficulty in finding an appropriate name for their product in such excellent resources as \url{greekmythology.com}, \url{theoi.com}, and wikipedia lists [\href{https://en.wikipedia.org/wiki/List_of_Greek_mythological_figures}{1}]
  [\href{https://en.wikipedia.org/wiki/List_of_minor_Greek_mythological_figures}{2}]
  [\href{https://en.wikipedia.org/wiki/List_of_Trojan_War_characters}{3}]. 
  
  Some major names yet to be taken, partially obtained by comparing the myth- and physics-themed indices below, are:
  from the primordial deities, 
  {\bf \em Erebus, Tartarus, Uranus};
  from the Titan race, 
  {\bf \em Calypso, Coeus, Crius, Helios, Iapetus, Mnemosyne, Oceanus, Phoebe, Rhea, Tethys, Themis};
  from the Olympian family,
  {\bf \em Aphrodite, Apollo, Ares, Artemis, Bacchus, Cora, Demeter, Dionysus, Hephaestus, Jupiter, Mars, Mercury, Neptune, Persephone, Poseidon, Vesta, Vulcan}; 
  from the mortals, 
  {\bf \em Aglaope, Ajax, Alcides, Asterion, Atalanta, Bellerophon, Cassandra, Clytemnestra, Chrysaor,  Electra, Eurydice, Hecuba, Helen, Jason, Jocasta, Laius, Menelaus, Minotaur, Narcissus, Odysseus, Paris, Peleus, Pelops, Perseus, Polyphemus, Priam, Telamon, Telemachus, Tiresias, Thelxiope}.\\

{\bf Other myth systems}   

  Why Greek myth? 
  That is best answered by the mythographer; it is not terribly plain to me why nomenclature in the sciences overwhelmingly favours one mythical frame. 
  But physics does occasionally help itself to names from non-Greek narratives.
  Thus (with no pretense to being exhaustive) Norse myth gives us \href{https://phys.org/news/2021-06-thor-collaboration-heavy-ion-collision.html}{\small THOR} and   \href{http://lhcb-comp.web.cern.ch/Analysis/loki/}{\small LOKI},
   Celtic lore \href{https://inspirehep.net/literature/361067}{\small TRISTAN}
 and \href{https://en.wikipedia.org/wiki/On-Line_Isotope_Mass_Separator}{\small ISOLDE},
   Hebrew mythology \href{https://en.wikipedia.org/wiki/MATHUSLA}{\small MATHUSLA} and \href{https://inspirehep.net/literature/515725}{\small NOE}, and
   {\em The Lord of the Rings} the \href{https://www.ing.iac.es/PR/SH/SH2007/sauron.html}{\small SAURON project} and the \href{https://conference-elbereth.obspm.fr/?Welcome-7}{Elbereth conference}. 
To my knowledge nothing yet in fundamental physics takes its name from the Hindu myths on which I grew up. 

\clearpage

\twocolumn

\section*{Acknowledgments}

I owe thanks to the following for new entries in the 3rd edition:
Marianne Moore (Zeus21),
and in the 2nd edition:
David d'Enterria (AURIGA, DELPHES, ELENA, PHENIX, SATURN),
Gabriel Orebi Gann (EOS [the detector]),
Andrew Larkoski (ARGUS),
Lina Necib (Nyx),
Dennis V. Perepelitsa (PHOBOS),
Albert Petrov (DAFNE),
Marcos Santander (PLUTO),
Kostantinos Tassis (PASIPHAE), and
Shawn Westerdale (Urania [the extraction plant] and Aria).
Peter Morse brought to my attention the \href{https://journals.aps.org/prl/abstract/10.1103/PhysRevLett.128.018002}{Apollonian and Dionysian packing schemes} of hard spheres. 
Corrections and suggestions by 
Adam Aurisano, 
Matheus Hostert,
Patrick Koppenburg,
Rafael Lang
and
Pranava Teja
have been valuable.

\section*{Further reading}

\href{http://nirmalraj.wikidot.com/miscellaneous:planetmythos}{\em The Wandering Tribe}, my retelling of the mythical narrative that supplied the names of Solar System bodies. 

-----

\section*{Bibiliography}

\ \ \ \href{https://www.goodreads.com/book/show/23522.Mythology}{\em Mythology} (1942), Edith Hamilton,

\href{https://www.goodreads.com/book/show/271385.Words_from_the_Myths}{\em Words from the Myths} (1961), Isaac Asimov,

\href{https://www.goodreads.com/en/book/show/759718.The_Adventures_of_Ulysses}{\em The Adventures of Ulysses} (1969), Bernard Evslin,

\href{https://www.goodreads.com/book/show/18404819-gods-demigods-and-demons}{\em Gods, Demigods and Demons} (1975), Bernard Evslin,

\href{https://www.goodreads.com/book/show/7989499-the-odyssey}{\em The Odyssey: a Graphic Novel} (2010), Gareth Hinds,

\href{https://www.goodreads.com/book/show/35074096-mythos}{\em Mythos} (2017), Stephen Fry,

\href{https://www.goodreads.com/book/show/41433634-heroes}{\em Heroes} (2018), Stephen Fry,

\href{https://www.goodreads.com/book/show/53443339-troy}{\em Troy} (2019), Stephen Fry.

\printindex
\printindex[indphy]

\end{document}